\documentstyle[11pt,IAU207_pasp,twoside]{article}
\input psfig

\def\edcomment#1{\iffalse\marginpar{\raggedright\sl#1\/}\else\relax\fi}
\reversemarginpar

\begin{document}
\title{Super Stellar Clusters in HII Galaxies}
 \author{Eduardo Telles}
\affil{Observat\'orio Nacional, Rua Jos\'e Cristino, 77 \\ 20921-400 - Rio de Janeiro - Brasil}

%\author{Ima Co-Author}
%\affil{The Name of My Institution, The Full Address of My Institution}

\begin{abstract}
{\small Stellar Clusters are identified in
images and in the spectra of these star forming dwarf galaxies.  These
Stellar Clusters have properties similar to those observed in other
violent star forming galaxies and may the elementary entities of a
starburst.}
\end{abstract}

\section{Introduction}
\def\msun{\thinspace\hbox{$\hbox{M}_{\odot}$}}
\def\etal   {et\nobreak\ al.\ }
\def\Zsun{\thinspace\hbox{$\hbox{Z}_{\odot}$}}

Super Star Clusters (SSC) are found not only in the starburst regions
of strongly interacting giant galaxies but also in dwarf galaxies in a
bursting phase of star formation (SF), HII galaxies.  The formation
and evolution of these Stellar Clusters may be an important new
ingredient in regulating the history of star formation in HII
galaxies, including the triggering mechanism responsible for
initiating this intense current episode of SF.  Since their discovery
in the early 70's, HII galaxies have gained the status of being the
nearest galaxies with properties expected for truly young galaxies in
the local universe.  The question, posed by Sargent \& Searle (1970)
of whether these are young galaxies forming stars for the first time
seems now to have been answered by recent studies.  H {\sc II}
galaxies show an underlying stellar population of intermediate to old
age (Telles \& Terlevich 1997, Doublier \etal 1997, Marlowe \etal
1999, Cair\'os \etal 2001).  Telles, Melnick \& Terlevich (1997) have
classified a large sample of HII galaxies into two main groups: Type I
includes the most luminous H {\sc II} galaxies, all of which present
an overall irregular morphology, and in many cases signs of being
merger products. The low luminosity Type II's, on the other hand, are
more compact and rounder showing no evidence of irregular isophotes or
any sign of being disturbed by external agents.  Type II HII galaxies
are the ones which mostly challenge our understanding of how SF was
initiated.

In this contribution I will review some of our own recent work, in
which my collaborators and I have been involved, in the study of star
formation in HII galaxies.  It is not my intention to review other
recent work in this area, but I point the reader to the review by
Kunth \& Ostlin (2000).

\section{What Triggers the Starburst in HII Galaxies?}
\begin{figure*}
\centerline{\hbox{
\psfig{figure=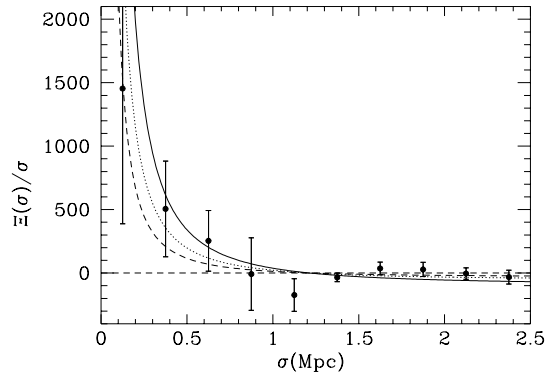,width=8.5cm,angle=270}
}}
\caption{{\small The projected cross-correlation between the HII galaxies and
the faint APM field galaxies, $\Xi_{hg}(\sigma)/\sigma$. The points
show our measurements. The solid, dotted and dashed lines show the predicted
$\Xi$ for $r_0=5.1, 3.7$ and $2.7 h^{-1}$ Mpc respectively. From
Telles \& Maddox (2000).}}
\label{fig:apm} 
\end{figure*}

Early studies of the environmental or clustering properties of HII
galaxies have agreed that these are isolated galaxies with a deficit
of $L>L^*$ galaxies within 1 Mpc (Campos-Aguilar \etal 1993, Vilchez
1995, Telles \& Terlevich, 1995).  These compelling studies have shown
that HII galaxies are not associated with giant galaxies, therefore
they are not tidal debris of strongly interacting systems.  These
results show there is no evidence for external triggers being
responsible for the onset of the current violent star formation
episode.

An alternative was first proposed by Melnick (1987) that HII galaxies
would be triggered not by giant galaxies but by other dwarfs or
intergalactic HI clouds.  Taylor and collaborators followed up these
ideas and using the VLA detected 12 HI companions around 21 HII
galaxies, while only 4 HI-rich companions were detected around a
control sample of 17 quiescent low surface brightness dwarfs (LSBD)
(Taylor 1997 and references therein).  Despite the effort by this
group, these results were not able to, beyond reasonable doubt, settle
the question of the possible effect of low mass companions as external
triggers of the SF in HII galaxies, because some HII galaxies do not
show HI companions while some LSBD do show HI companions and these are
{\em not} 'bursting'!

Telles \& Maddox (2000) went further, in an attempt to supersede these
small number statistics and put further constraints on the existence
and/or properties of possible low mass companions. We carried out an
investigation of the galaxy environments of a sample of over 160 low
redshift HII galaxies by cross-correlating their accurate position in
the sky to faint field galaxies in the APM galaxy catalog.  By using
the fact that all detectable HI clouds in 21 cm surveys have an
optical counterpart, clouds with masses greater than $10^8$\msun would
be detected in our statistics analysis, with reasonable assumptions
regarding their M(HI)/L$_B$ ratios.  The main result of Telles \&
Maddox can be summarized in Figure~\ref{fig:apm}.  The projected
cross-correlation measurements are marginally lower than the
predictions expected for a sample of normally clustered galaxies, and
lie between the auto-correlation functions of normal galaxies and HII
galaxies. The lower amplitude emphasizes, that there is not a large
excess of near neighbors around the HII galaxies compared to normal
galaxies.  Again this is incompatible with suggestion that HII
galaxies are triggered by tidal interactions with nearby low-mass
galaxies.

These results put further limits on the properties of the possible low
mass candidate triggers of star formation in HII galaxies.  These may
be low mass companions at distances lower than about one or two
hundred pc, otherwise we would have picked them up in the correlation
function analysis.  Pustilnik \etal (2001), in another statistical
analysis of the environment of 86 low mass galaxies with active star
formation from the Second Byurakan Survey, find indication for close
companions that might have been missed out in our cross-correlation
function.  In any case, it has become clear that some (rather large)
fraction of HII galaxies are indeed isolated.

The point here is not only the presence of candidate triggers but the
mechanism of star formation triggering.  It has been shown in past
studies that star formation is enhanced by galaxy interactions, with
clear evidence in large mergers, as in the high luminosity, high mass
end of the starburst phenomena (e.g. the ultra-luminous IRAS galaxies);
and other examples of strong tidal effects.  The fact that HII
galaxies are dominated by a large emission line region has inspired
the hypothesis that one has to have a very strong tidal agent to make
all the gas  rapidly pile up in the center of the galaxy so that the
burst would be seen as one large event. The observations of structures
in the ISM and the presence of different knots of SF, may free us from
the idea of a large single event but indicate that SF may have formed
locally and propagated.  For this a weak tidal force may suffice, such
as the mechanism proposed by Icke (1985). Shocks, due to internal or
external agents, causing supersonic motions to propagate, producing
high pressures and high density clouds enable the formation of the
stellar clusters. Once they are formed their evolution may regulate
further propagation of star formation.  Evidence for the presence of
stellar clusters in starbursts is growing and one may ask whether
they are the cause or the effect of the starburst and in what
conditions it may be one case or the other.

\section{Structure and Supersonic Motions in HII Galaxies}

\begin{figure*}[ht]
\centerline{\hbox{
\psfig{figure=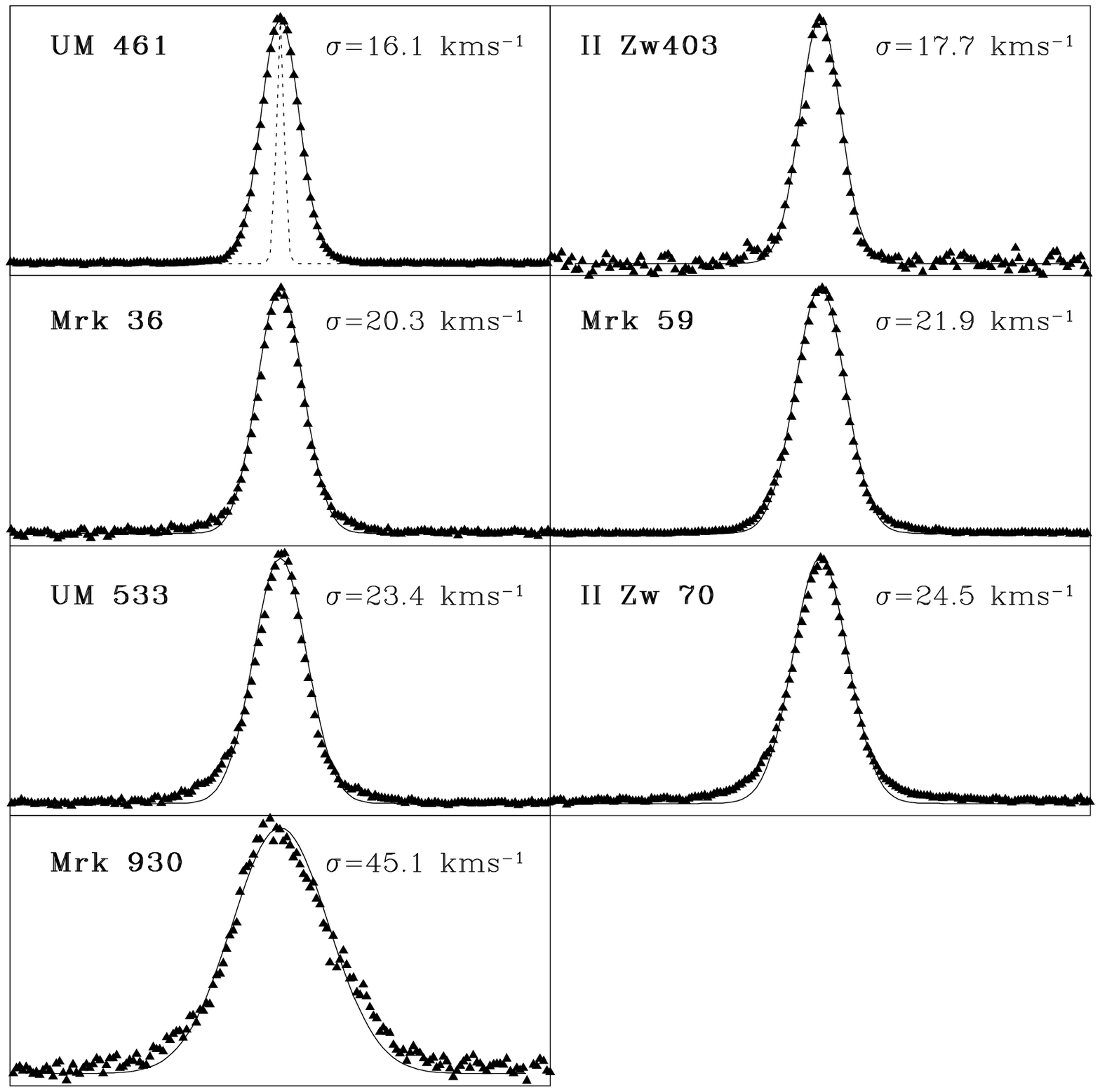,width=6.5cm,angle=0}
\psfig{figure=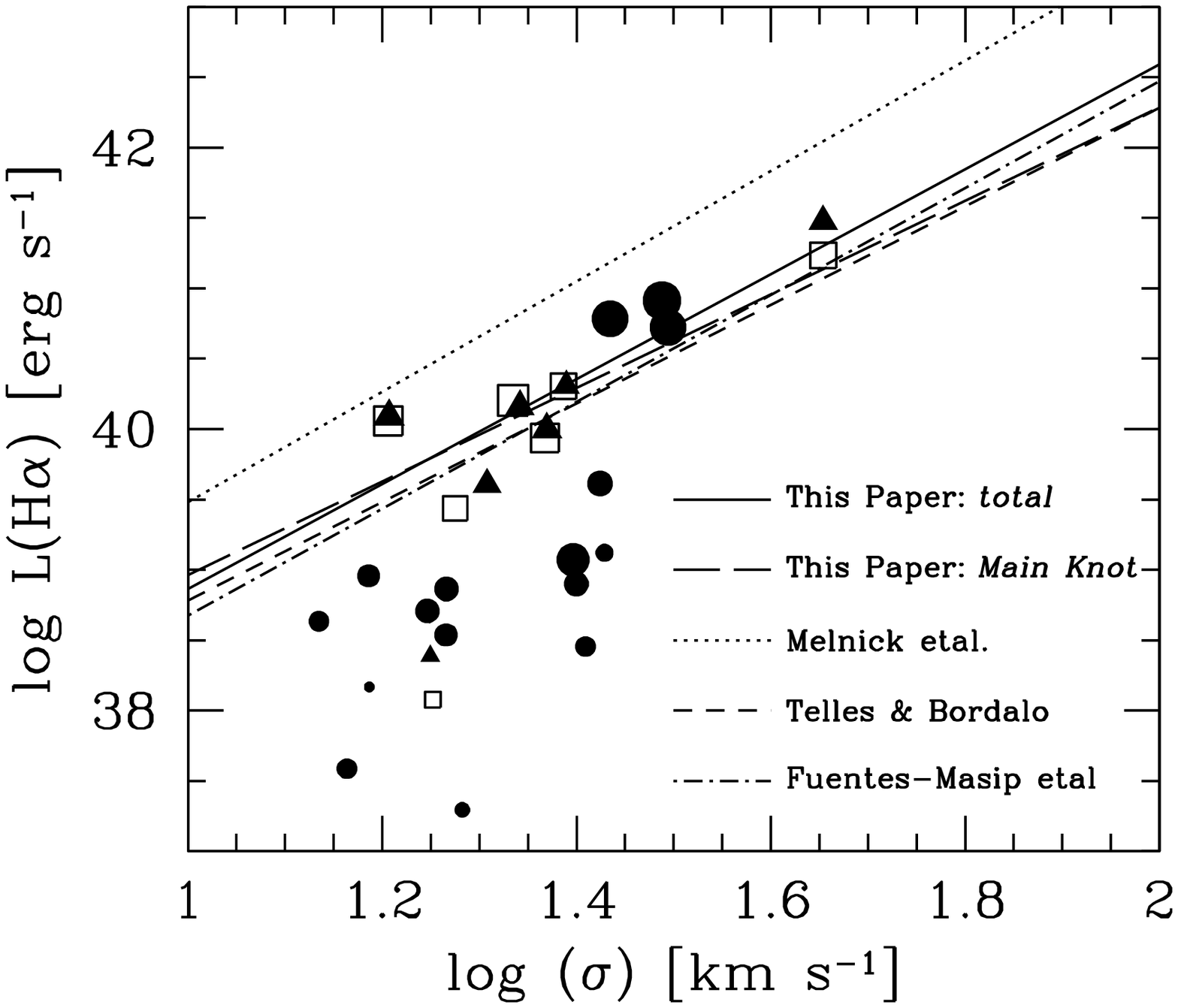,width=7.5cm,angle=0}
}}
\caption{{\small a. Emission line profiles of HII galaxies on the H$\alpha$
line. Also shown are the Gaussian fits used to derive the final gas
velocity widths.  b. The luminosity - line width relation. Points:
{\it total} apertures (solid triangles), Main Knot (open squares),
other regions or knots (solid circles).  Point sizes represent
relative H$\alpha$ surface brightness. The lines are simple linear
square fits for our data and the statistical works from the
literature. }}
\label{fig:wht}
\end{figure*}

Our high resolution echelle observations of a sample of HII galaxies
(Telles, M\'unoz-T\'u\~non \& Tenorio-Tagle 2001) showed a strong
variation in the line profiles across the emitting region and even in
the most compact sources, there is an indication of separate knots of
star formation (SF) evolving concurrently within the galaxy nucleus.
The presence of multiple knots of star formation within the line
emitting regions is also observed on high spatial resolution images,
in particular in the near-IR, where we also identified knots which are
possible super stellar clusters (SSC). 

The structure found in H {\sc II} galaxies has profound implications
on several topics. In particular on issues such as star formation and
its possible sequential propagation in H {\sc II} galaxies, and how 
the ISM is structured in these galaxies. Another issue central in this
field of research is the validity of the interpretation, and use of
the empirical correlations, of size and luminosity {\it vs} their
supersonic line widths for high redshift galaxies.  These correlations
were first found for Giant H {\sc II} Regions by
Terlevich \& Melnick (1981) and Melnick \etal (1987) and extended to H
{\sc II} galaxies by Melnick \etal (1988) and Telles \& Terlevich
(1993).  The similarity of these relations with the Faber-Jackson
relation for elliptical galaxies and the Tully-Fisher for late type
spirals have led these investigators to propose that gravity is the
dominant mechanism ruling the overall dynamics of HII galaxies. Other
competing theories invoke stellar evolution with massive star winds
and supernovae to be responsible for the inner motions in HII
galaxies.  However, the fact that line profiles are observed to be
well fitted by gaussians (Figure~\ref{fig:wht}a.) indicate that a
continuous source of energy input is required.  In addition, the
[L-R-$\sigma$] correlations (Figure~\ref{fig:wht}b.) find a natural
explanation in the Virial Theorem, which makes the gravity model the
most appealing theory for the motions in HII galaxies.

Details of this study can be found in our paper (Telles \etal 2001)
but our main points can be summarized: (i) our results confirm and
extend the empirical correlations found for giant H {\sc II} regions
and H {\sc II} galaxies.  We indicate the possible effect of a second
parameter in the relations similar to the fundamental plane for
elliptical galaxies.  (ii) enhanced spectral and spatial resolution
seems to unveil an intricate structure in H {\sc II} galaxies.  (iii)
H {\sc II} galaxies when resolved, present several emitting knots with
a variety of shapes, luminosity and $\sigma$ values.  (iv) The
intrinsic properties (luminosity, velocity dispersion) of a galaxy are
dominated by the central (core) component.

A fine calibration of these relations for local H {\sc II} galaxies
may be of great importance if used as a distance indicator of galaxies
at large redshift, since H {\sc II} galaxies are easy to find at great
distances (Melnick, Terlevich \& Terlevich 2000).

\section{The History of Star Formation through their Stellar Clusters}

\begin{figure*}
\centerline{\hbox{
\psfig{figure=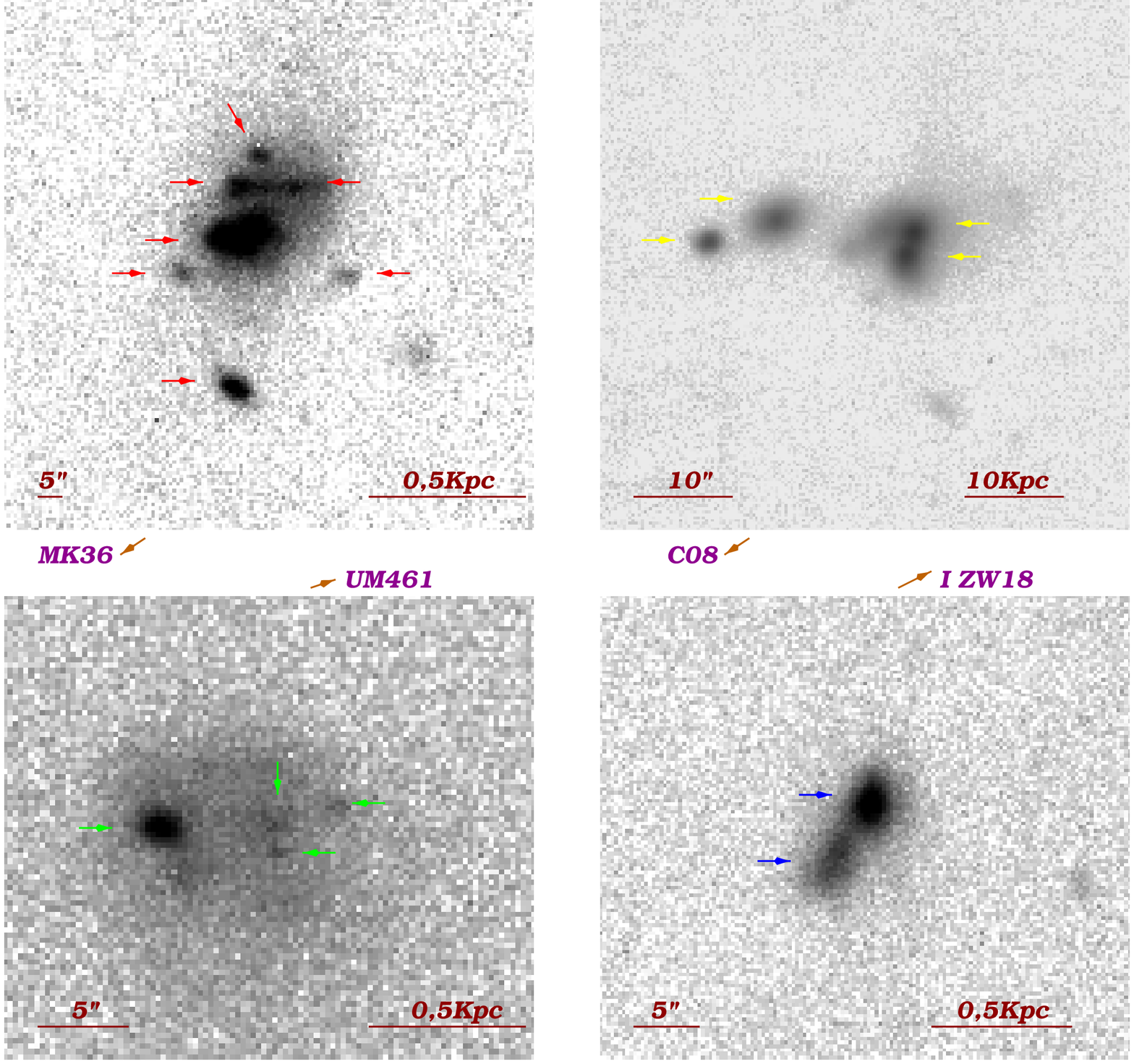,width=7cm,angle=270}
\psfig{figure=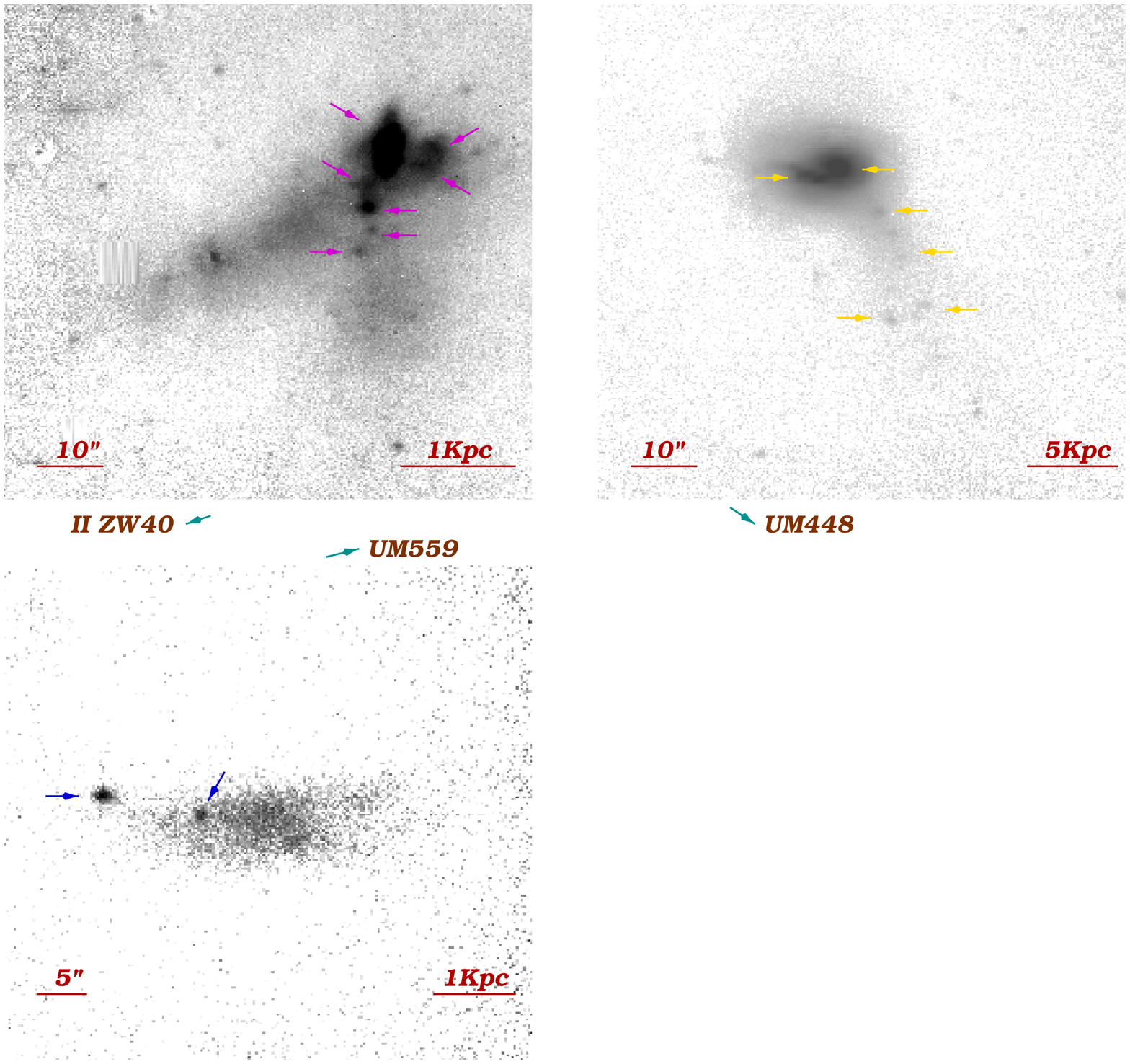,width=7.3cm,angle=270}
}}
\caption{{\small J band images of our UKIRT sample of HII galaxies.
Arrows identify some of the SSC detected.}}
\label{fig:images}
\end{figure*}

Our near-IR images of H {\sc ii} galaxies have revealed
supercluster-sized objects (SSC) within the star forming regions
similar to those initially detected in HST ultraviolet images of
starburst galaxies (Meurer et al.\ 1995; Vacca 1994, Ostlin \etal
1998).  In many cases these knots form continuous structures, which
suggests that star formation may have propagated across the starburst
region.  These SSCs appear to be a common feature in the starburst
regions of merging galaxies (Surace \& Sanders 1999, see also
Whitmore's and O'Connell's contribution), and have also been
identified in a sample of Wolf-Rayet galaxies in ground based optical
observations by Mendez \& Esteban (2000).  Our present study, which
will be appearing in a forthcoming paper in the near future (Telles,
Sampson \& Tapia, 2001, in preparation), have shown that these SSC are
more clearly identified in the near-infrared.  Figure~\ref{fig:images}
shows J band images of our sample observed at UKIRT with the
identification of some of the most visible clusters.  We have
attempted to age date the individual star clusters, thus having an
alternative method of determining the star formation history of the
galaxy.  This seems to do a better job when comparing with models of
population synthesis since we disentangle the need to assume a history
of star formation a priori to age date the galaxy from its integrated
light because the assumption of an instantenous burst for each
individual stellar cluster is better justified.

We have used the Starburst99 model (Leitherer \etal 1999) of population
synthesis to compare with our observations.  Many words of caution
must be given at this stage but I will defer these to the paper and
only cite that one must worry about the choices of model parameters
such as IMF, extinction, metallicity, nebular emission and dust
emission.  Due to the wide choices of model parameters and
assumptions, it may not be possible or useful to derive the absolute
ages from these comparisons, but the relative ages of the SSC from one
single galaxy may be useful to describe the temporal and spatial
patterns of Star Formation.

\subsection{Temporal and Spatial properties of Star Formation}

\begin{figure*}
\centerline{\hbox{
\psfig{figure=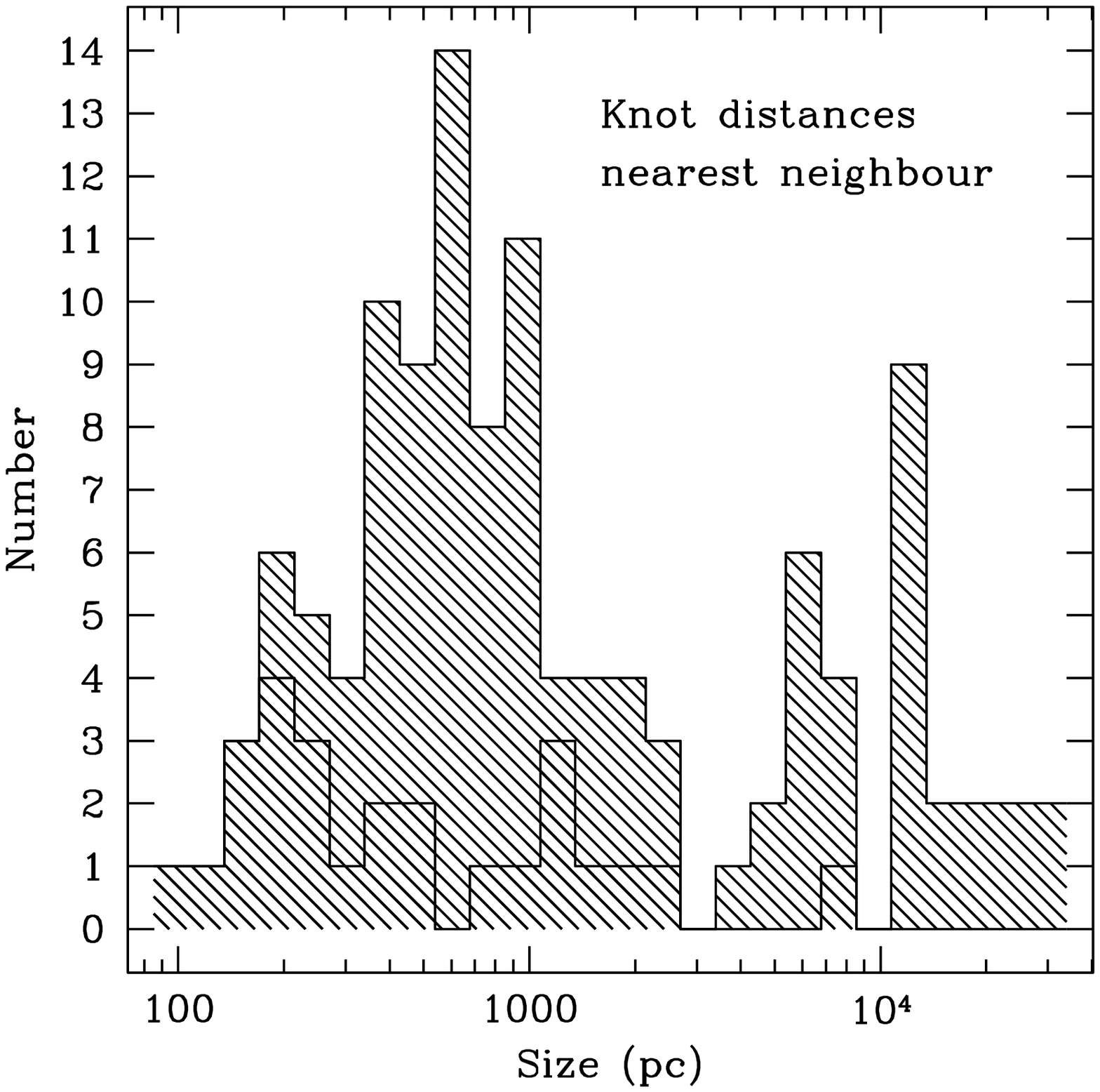,width=7cm,angle=0}
\psfig{figure=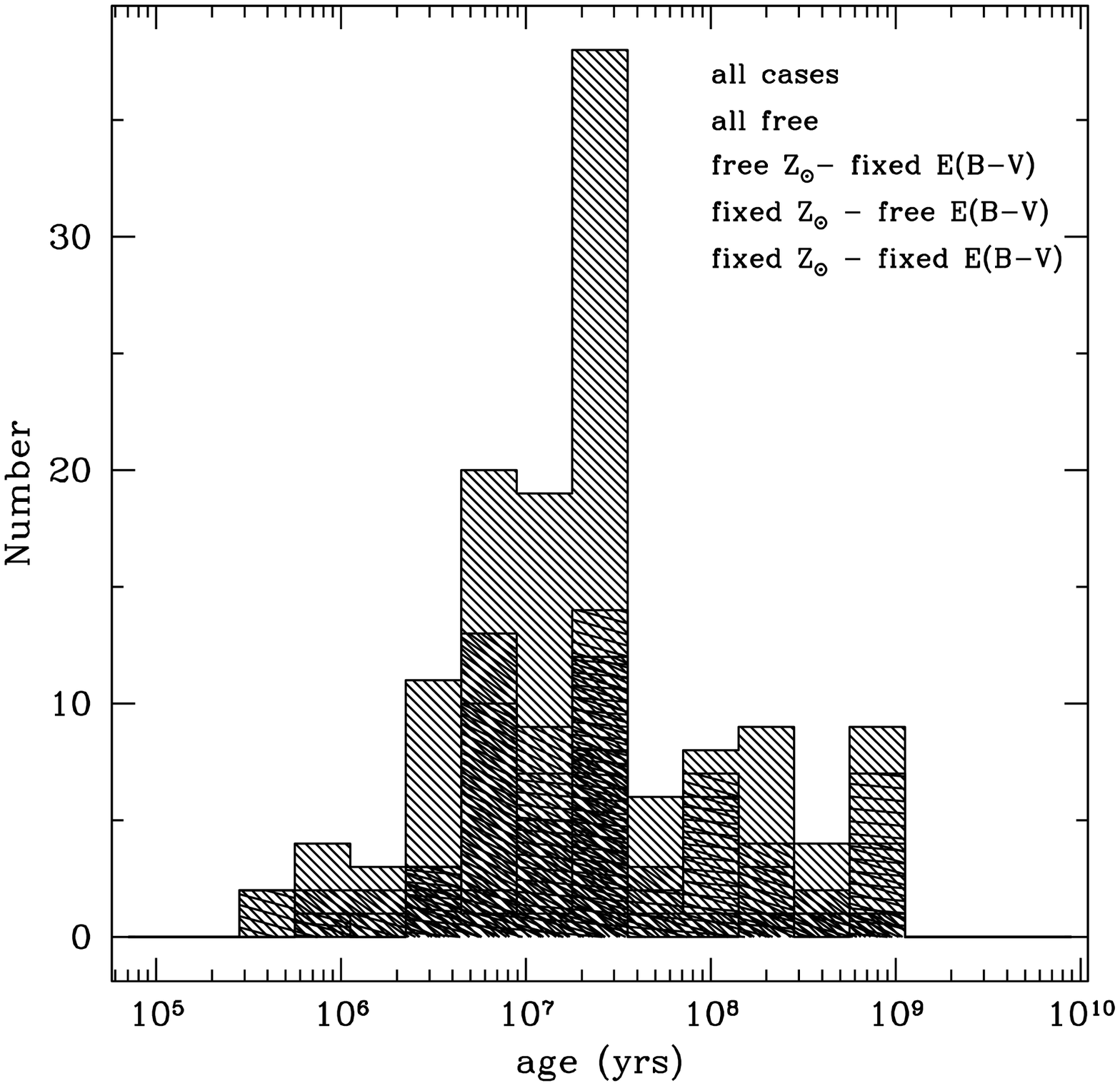,width=7cm,angle=0}
}}
\caption{{\small a. Distribution of distances from cluster to cluster.  Inner
histogram represents the nearest neighbors(H$_o$ = 65 km s$^{-1}$
M$^{-1}$.). b. Distribution of derived ages. Details are given in our
paper. Only the overall results are intended to be noted here. }}
\label{fig:dists}
\end{figure*}

We have combined the information of our observations in the optical V
R I and near-infrared J H K bands to compare with the model
predictions.  Our preferred choice of model metallicity is the one
derived from spectroscopy and assumed to be an upper limit to the
metallicity of the stellar population. Our preferred choice of
extinction is also derived from spectroscopy measured by the Balmer
decrement (H$\alpha$/H$\beta$ ratio) and assumed to be also an upper
limit to the extinction in the continuum (E(B-V)$_{cont}= 0.44 \times$
E(B-V)$_{gas}$) as given by Calzetti \etal (2000).  Despite the
preferred (fixed) values of metallicity or extinction, we ran the
comparison with the models of all metallicities and let the extinction
also be a free parameter in order to verify which best represented all
our combined observations.  I will present here the overall results
for all galaxies and super stellar cluster combined and refer the
results for individual clusters and galaxies to the paper.

Figure~\ref{fig:dists}a shows the histogram of all distances of a SSC
to another in  single galaxies for all galaxies.  The inner histogram
shows the nearest neighbors only.  It can be seen that typically the
nearest neighbor distance is about a few hundred pc.
Figure~\ref{fig:dists}b  shows a histogram of the derived ages of the
individual clusters.  I will not mention details of these analyses
here but results seem to indicate typical ages of a few 10$^7$ yrs.

Therefore we may speculate from this preliminary analysis that either
SF propagates very rapidly ($< 10^8$ yrs) or it is synchronized in
scales of hundreds of parsec so that a large burst can be observed to
dominate the whole extent of an HII galaxy.  Star Formation
propagation has been proposed and studied since the late 70's (Gerola
\& Seiden 1978; see also Nomura \& Kamaya 2001).  We have also learned
a great deal from the properties observed in the best laboratory for
the study of massive star formation in a sub-solar metallicity
environment, the giant HII region 30 Dor in the LMC. Grebel \& Chu
(2000) show how SF may have propagated in this smaller analog to HII
galaxies.

\begin{figure*}
\centerline{\hbox{
\psfig{figure=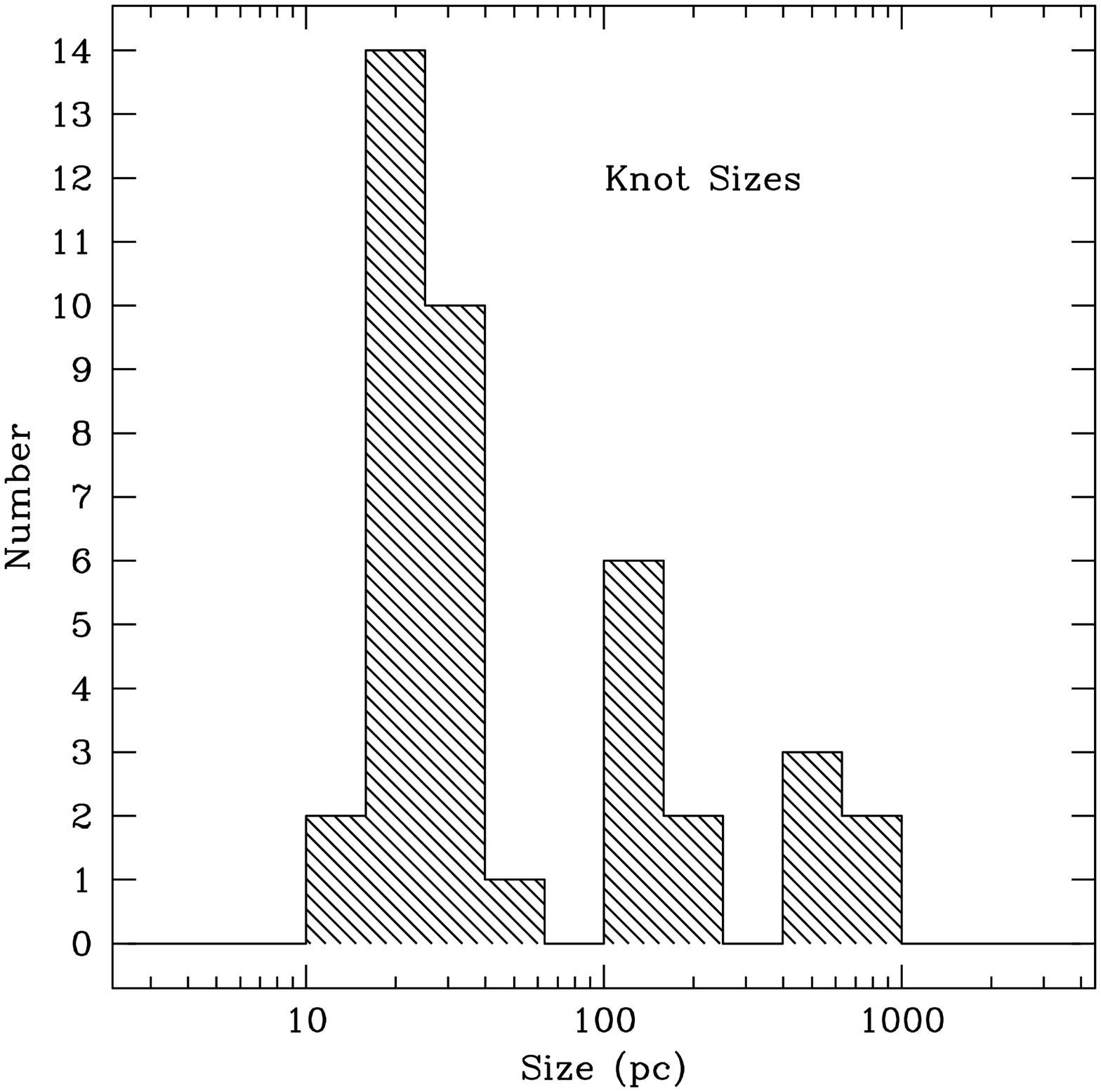,height=5.5cm,angle=0}
\psfig{figure=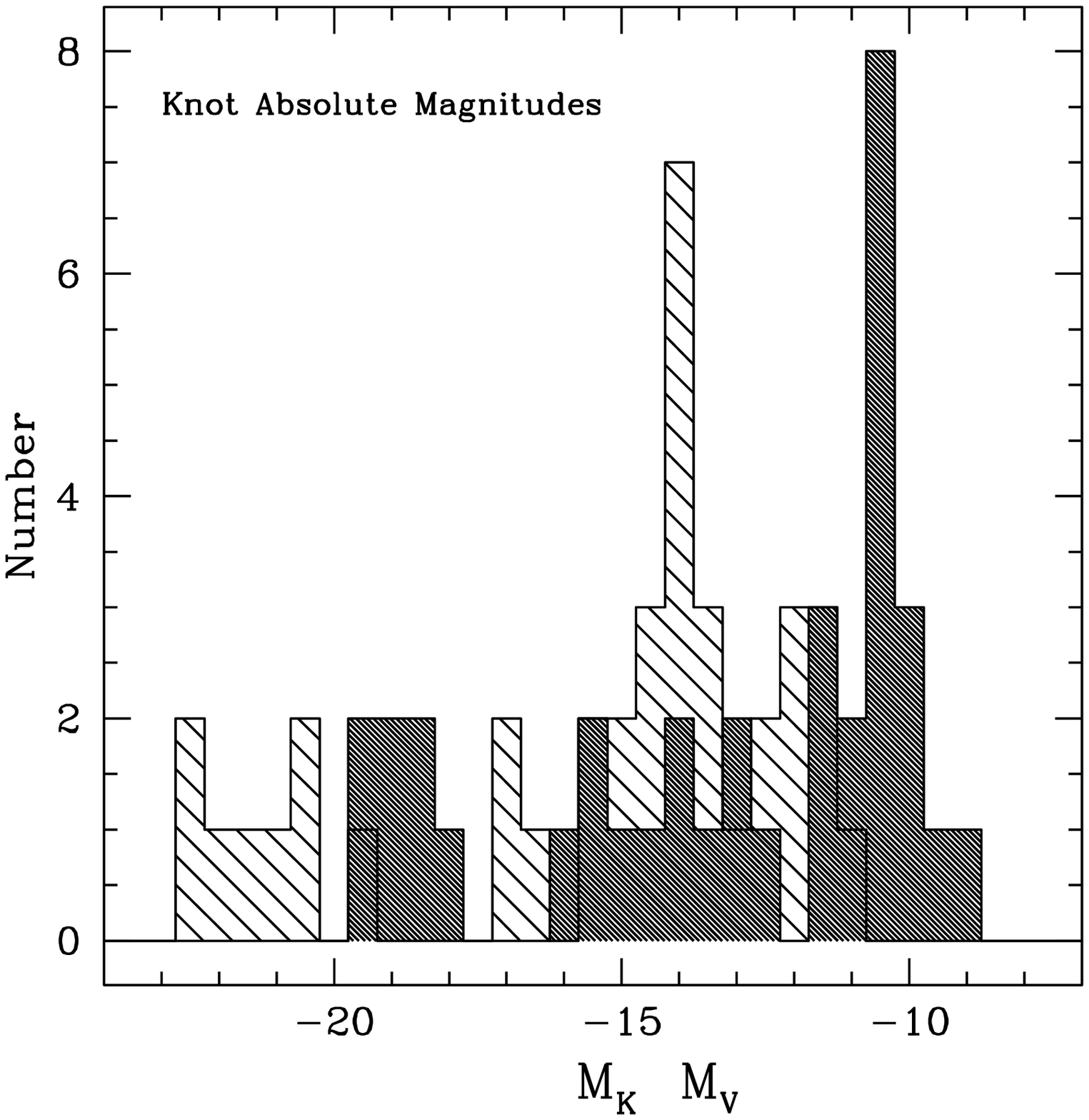,height=6cm,angle=0}
}}
\caption{{\small Distribution of Sizes and Absolute Magnitudes of SSCs.}}
\label{fig:sizes}
\end{figure*}

\subsection{Are these SSC similar to those observed in interacting galaxies?}

A first question one may pose is how these SSC in HII galaxies compare
with those in strongly interacting or merging giant systems observed
with the HST.  Figure~\ref{fig:sizes} shows the histograms of sizes
and absolute magnitudes of all SSC.  Disregarding the largest and most
luminous ones which represent the nuclear regions, and are actually an
ensemble of unresolved clusters, we find the SSC to have typical sizes
(R$_{eff}$ of about 20 pc, M$_V \approx -10$ and M$_K \approx -14$.
Assuming the results of the stellar models, these numbers represent a
stellar cluster of one million stars (1-100 \msun) with a Salpeter
IMF, not disimilar to the ones observed in other starbursts.  We may
have identified the elementary entities of which starbursts consist.
 
\subsection{Is there anything wrong with the comparison of models and observations of low metallicity galaxies?}

\small
\begin{center}
\begin{table}
\centering
\begin{tabular}{ccccccccc} \hline
\multicolumn{5}{c}{Starburst99}& &  & &  \\ \hline
a  &  b  & c & d  & e & &  & &  \\ \hline
2\Zsun & \Zsun & 2/5\Zsun & 1/5\Zsun & 1/20\Zsun & $<\log(age)>$ &$<$E(B-V)$>$     & {\footnotesize metallicity}& {\footnotesize extinction} \\ \hline
21 &  9  & 2 & 3  & 5 & 7.17+-0.65 & $0.24\pm0.25$ & {\bf free}  & {\bf free}  \\
14 &  13 & 3 & 4  & 6 & 7.30+-0.57 & $0.20\pm0.07$ & {\bf free}  & {\it  fixed} \\
0  &  0  & 7 & 24 & 9 & 7.52+-0.86 & $0.37\pm0.26$ & {\it  fixed} & {\bf free}  \\
0  &  0  & 7 & 24 & 9 & 7.64+-0.76 & $0.20\pm0.07$ & {\it  fixed} & {\it  fixed} \\ \hline
\end{tabular}
\caption{{\small Best results for our comparison of Starburst99 model with our combined VRIJHK colors for 
different metallicities and extinction}}
\label{tab:res}
\end{table}
\end{center}

\normalsize

A first analysis of our overall results is summarized in
Table~\ref{tab:res}. The first 5 columns represent the number of SSC
for which that particular metallicity has proven to be the best
result.  The last two columns tell you whether these results were
obtained for a preferred chosen metallicity ({\it fixed}) or left as a
free parameter ({\bf free}). Columns 6 and 7 give the corresponding
mean age or extinction with their respective rms (not an error).

If we take these results at face value, despite all other
complications, one can say that the models at low metallicity are not
as red as they should be to reproduce the observed colors, because the
best results are given by high metallicity models (model a or b),
regardless of extinction (which is shown to vary within one galaxy),
while we know that these are low metallicity systems.  A K band excess
has also been pointed out by Leo Vanzi (this conference) in his
analysis of SBS0335-052. Therefore, these best results cannot be right.

The problem may be caused by a number of things: i) We are using the
wrong models; we cannot for instance detect clusters with ages greater
than 1 Gyr. We have tested this hypothesis by using Bruzual \& Charlot
(2000) models but our present conclusions do not seem to change. ii)
there may be an important contribution from dust emission.  Although
this has to be tested by observations at longer wavelengths, it is
unlikely because emission line ratios do not point that way. iii) most
likely, models at low metallicity are failing to reproduce the
observations. One possible reason, among others, is that evolutionary
tracks at low metallicity do not seem to reproduce either the observed
numbers or colors of Red Supergiants (Maeder 2001, private
communication). Stellar models which include rotation are being
devised to produce a longer and more numerous RSG phase at low
metallicity.

\section{Some Conclusions}

\noindent
- Stellar Cluster formation and evolution is the mode of SF in HII
galaxies, and is responsible for the magnitude and efficiency of the
present burst. \\

\noindent
- The properties of SSC in HII galaxies are similar to the SSC in
interacting galaxies, and may be generic in a starburst.\\

\noindent
- HII galaxies are not triggered by strong tidal agents. Weak tidal
agents may be present at distances of only a few hundred parsec and
masses of only $\approx 10^8$\msun.  A significant fraction may not be
triggered at all and internal mechanism may be at play to let loose SF
and its propagation.\\

\acknowledgments 
I am deeply indebted to my collaborators Casiana Mu\~noz-Tu\~n\'on,
Guillermo Tenorio-Tagle, Jorge Melnick, Mauricio Tapia, Roberto
Terlevich and Steve Maddox. I also thank my students Vinicius Bordalo
and Leda Sampson for their hard work in dealing with the data and
figures.  Finally, I thank the organizers and the IAU for their
financial support and for such a fruitful meeting.


\begin{references}



\reference Calzetti, D. \etal, 2000, \apj, 533, 682
  
\reference Campos-Aguilar,A., Moles,M., Masegosa,J., 1993, \aj, 106, 1784  

\reference Cair\'os, L.M., Vilchez, J.M.,
Gonz\'alez--P\'erez, J.N., Iglesias--P\'aramo, J. and Caon, N. 2001,
\apjs, 133, 321

\reference  Doublier, V., Comte, G.,
Petrosian, A., Surace, C, Turatto, M., 1997, \aaps, 124, 405

\reference Gerola,H. \& Seiden,P.E., 1978, \apj, 223, 129

\reference Grebel,E.K. \& Chu, Y.-H, 2000, AJ,119,787

\reference Icke, V., 1985, \aap, 144,115

\reference Kunth, D.  \& Ostlin, G. 2000, A\&ARv, 10, 1

\reference Leitherer \etal 1999, \apjs., 123, 3

\reference Marlowe, A.T., Meurer, G.R. \&
Heckman, T.M., 1999, \apj, 522, 183

\reference Melnick,J., 1987, in {\it \,``Starburst and Galaxy Evolution\,''}, 
eds. T.X.Thuan, T.Montmerle \& J.Tran Thanh Van, editions Fronti\`eres Gif 
Sur Yvette, France, p. 215

\reference Melnick,J., Moles M., Terlevich
R. \& Garcia-Pelayo J.M., 1987, \mnras, 226, 849

\reference Melnick,J., Terlevich R. \& Moles
M., 1988, \mnras, 235, 297

\reference Melnick,J., Terlevich,R. \&
Terlevich,E., 2000, \mnras, 311, 629

\reference Meurer, G. R., Heckman, T. M., Leitherer, C., Kinney, A.,
Robert, C., \& Garnett, D. R. 1995, \aj, 110, 2665

\reference Mendez,D.I. \& Esteban, C., 2000, A\&A, 359:493

\reference Nomura, H. \& Kamaya, H., 2001, \apj, 121 1024

\reference Ostlin \etal 1998, \aap, 335, 850


\reference Pustilnik, S.A., Kniazev, A.Y., Lipovetsky, V.A. \&
Ugryumov, A. A., 2001, astro-ph/0104334

\reference Sargent W.L.W. \& Searle L., 1970, \apjl, 162, L155

\reference Surace, J. A. \& Sanders, D. B., 1999, \apj, 512, 162

\reference Taylor,C.L 1997, \apj, 480, 524

\reference Telles, E. \& Maddox, s. 2000, \mnras, 311,  307

\reference Telles,E. \& Terlevich,R., 1993, \apss, 205, 49

\reference Telles,E. \& Terlevich,R., 1995, \mnras, 275, 1

\reference Telles, E. \& Terlevich, R. 1997, \mnras, 286, 183
 
\reference Telles, E., Melnick,J. \& Terlevich,R. 1997, \mnras, 288, 78

\reference Telles,E., Casiana Mu\~noz-Tu\~n\'on \& Guillermo
Tenorio-Tagle, 2001, \apj, 548, 671

\reference Terlevich, R. \& Melnick, J., 1981, \mnras, 195, 839

\reference Vacca, W.D., 1994, in ``Violent Star Formation'', ed Tenorio-Tagle, p 297

\reference Vilchez,J.M., 1995, \aj, 110, 1090


\reference

\end{references}
\end{document}